\documentclass{vgtc}                          

\ifpdf
  \pdfoutput=1\relax                   
  \usepackage{graphicx}                
  \DeclareGraphicsExtensions{.pdf,.png,.jpg,.jpeg} 
\else
  \usepackage{graphicx}                
  \DeclareGraphicsExtensions{.eps}     
\fi%

\graphicspath{{figures/}{pictures/}{images/}{./}} 

\usepackage{microtype}                 
\PassOptionsToPackage{warn}{textcomp}  
\usepackage{textcomp}                  
\usepackage{mathptmx}                  
\usepackage{times}                     
\usepackage{cite}                      
\usepackage{tabu}                      
\usepackage{booktabs}                  
\usepackage{balance}

\onlineid{1089}

\vgtccategory{Research}

\vgtcinsertpkg

\def\etal{\emph{et~al}.}

\usepackage{amsmath}
\usepackage{amssymb}
\usepackage[capitalise]{cleveref}
\usepackage{tikz}
\usepackage{balance}
\usepackage{booktabs}

\title{Visual Analysis of Large Multi-Field AMR Data on GPUs Using\\Interactive Volume Lines}

\author{Stefan Zellmann\thanks{e-mail: zellmann@uni-koeln.de}\\ %
        \scriptsize University of Cologne %
\and Serkan Demirci\thanks{e-mail: serkan.demirci@bilkent.edu.tr}\\ 
      \scriptsize Bilkent University %
\and U\u{g}ur G\"{u}d\"{u}kbay\thanks{e-mail: gudukbay@cs.bilkent.edu.tr}\\ 
      \scriptsize Bilkent University %
}

\teaser{
 \centering
\vspace{-2em}
\begin{tikzpicture}
 \node[anchor=south west,inner sep=0] (image) at (-5.5,0) {\includegraphics[width=0.245\linewidth]{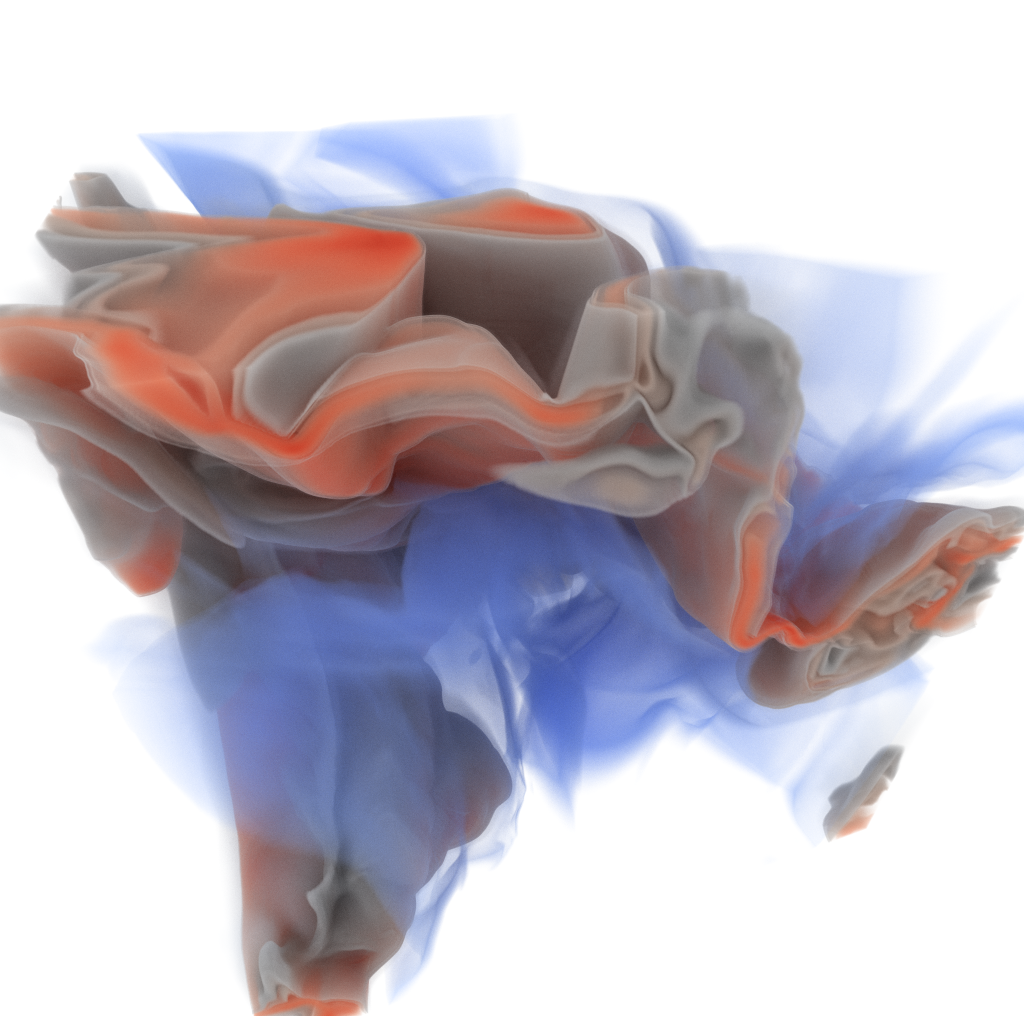}};
 \node[anchor=south west,inner sep=0] (image) at (-1.45,0) {\includegraphics[width=0.245\linewidth]{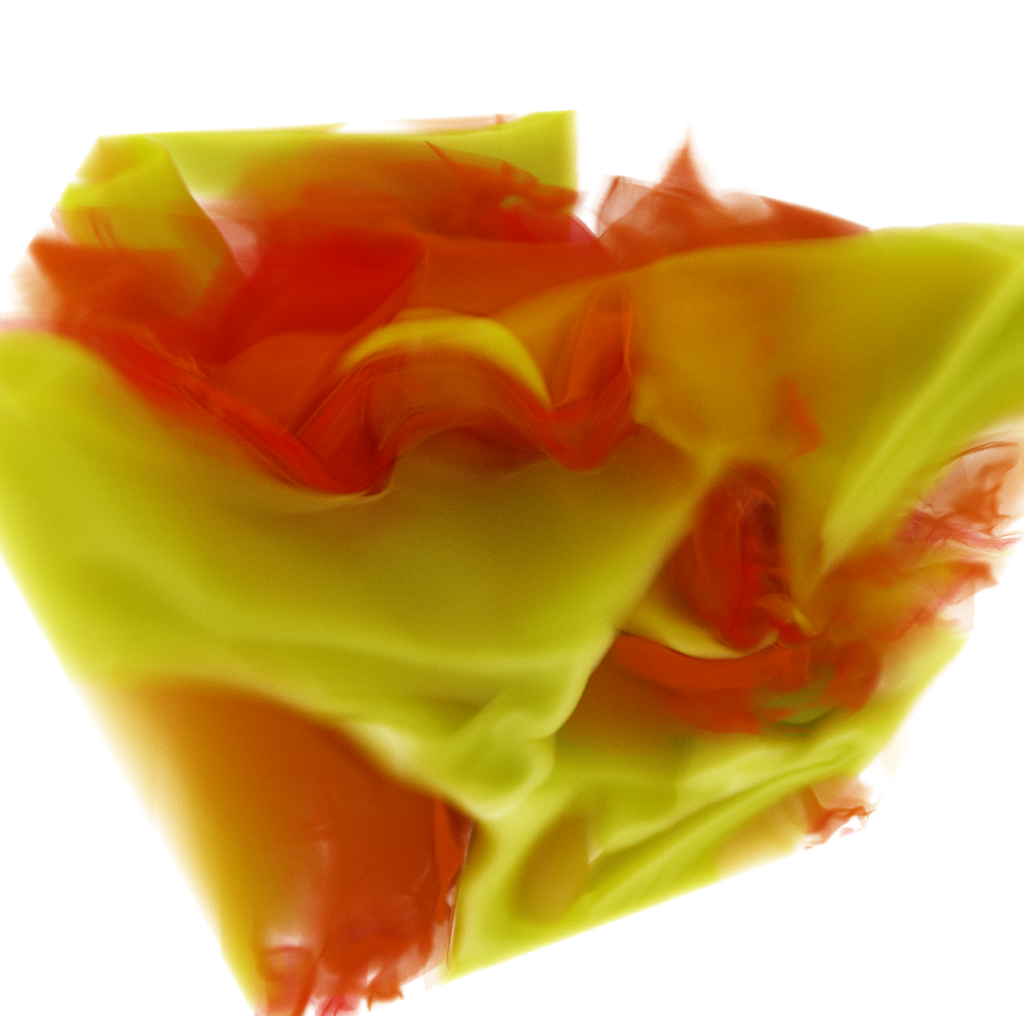}};
 \node[anchor=south west,inner sep=0] (image) at (2.6,0) {\includegraphics[width=0.245\linewidth]{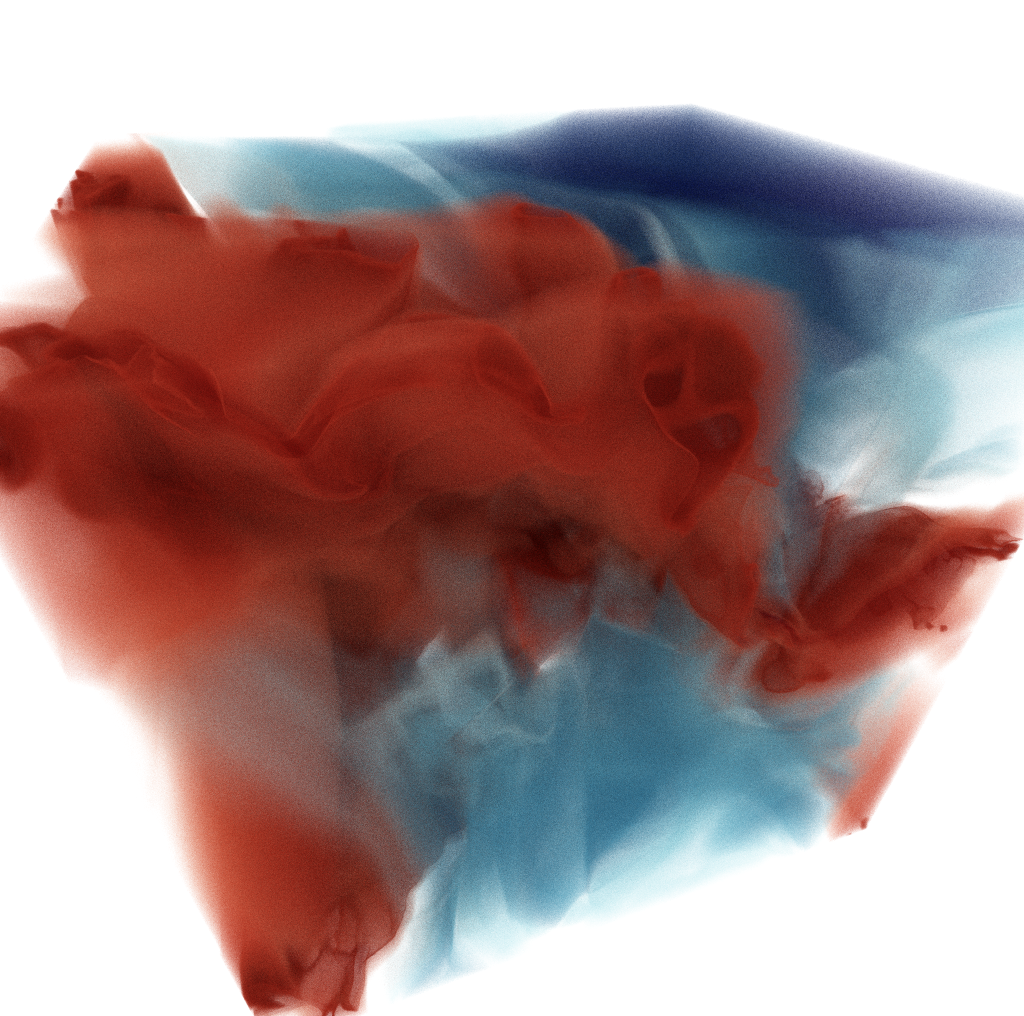}};
 \node[anchor=south west,inner sep=0] (image) at (6.65,0) {\includegraphics[width=0.245\linewidth]{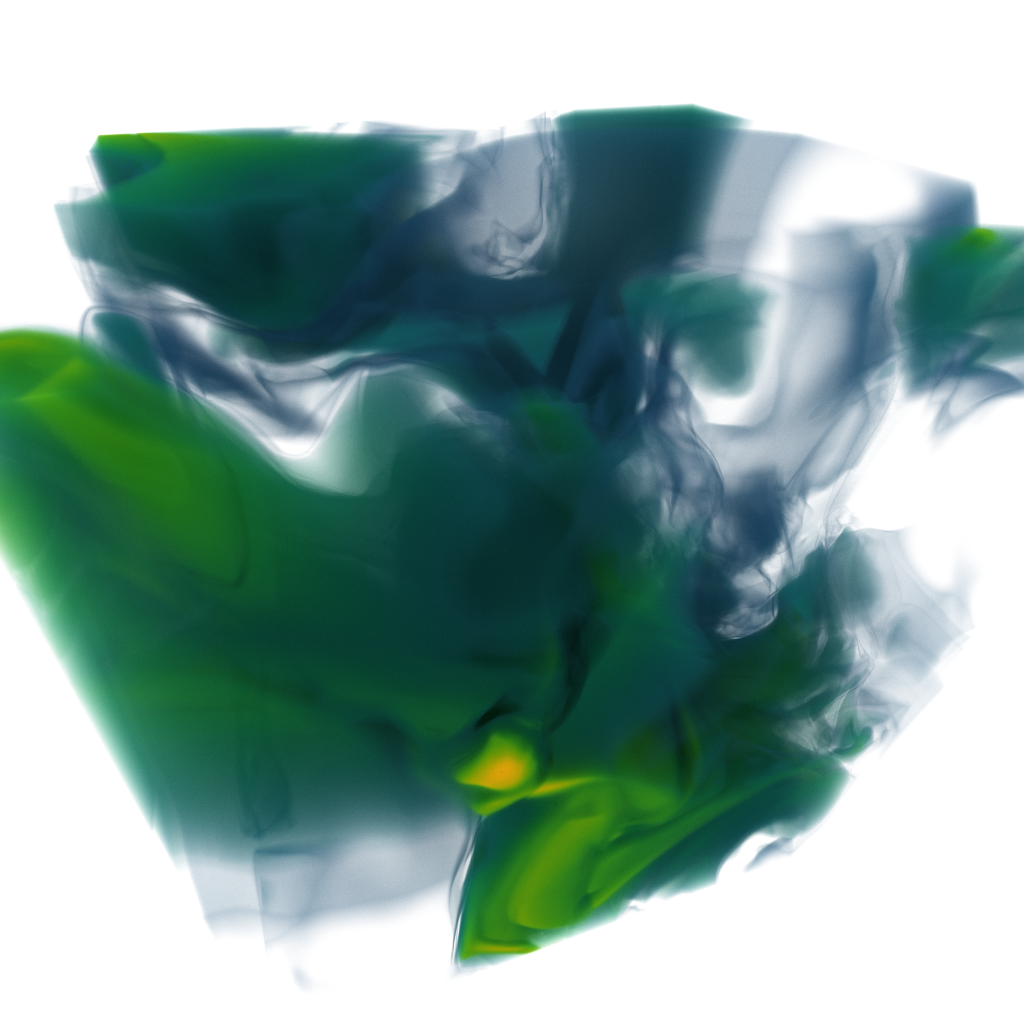}};

 \node[anchor=south west,inner sep=0] (image) at (-5.5,-1) {\includegraphics[width=0.999\linewidth]{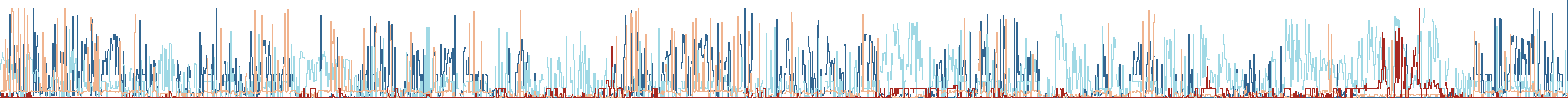}};
 \node[anchor=south west,inner sep=0] (image) at (-5.5,-2.03) {\includegraphics[width=0.999\linewidth]{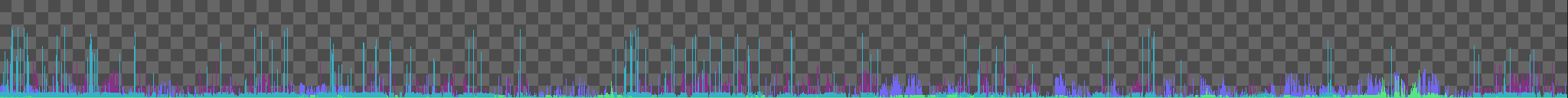}};

 \node[shape=circle,draw,fill=white,minimum size=14pt,inner sep=0, line width=1pt] (char) at (-5.2,3.4) {a};
 \node[shape=circle,draw,fill=white,minimum size=14pt,inner sep=0, line width=1pt] (char) at (-5.2,-0.27) {b};
 \node[shape=circle,draw,fill=white,minimum size=14pt,inner sep=0, line width=1pt] (char) at (-5.2,-1.30) {c};
\end{tikzpicture}

\vspace{-1em}
\caption{\label{fig:teaser}%
Coupled \emph{interactive volume lines} (IVL) and large volume visualization.
(a)~shows 3D renderings of four fields of an astrophysical simulation (from
left to right: density, temperature, pressure, and velocity magnitude). Our
prototype interactively links such large-scale AMR data with 1D volume line
plots (polyline plot representation in~(b), and bar plot representation
in~(c)).}
\vspace{-0.5em}
}
\abstract{To visually compare ensembles of volumes, \emph{dynamic volume lines}
(DVLs) represent each ensemble member as a 1D polyline. To compute these, the
volume cells are sorted on a space-filling curve and scaled by the ensemble's
local variation. The resulting 1D plot can augment or serve as an alternative
to a 3D volume visualization free of visual clutter and occlusion.
Interactively computing DVLs is challenging when the data is large, and
the volume grid is not structured/regular, as is often the case with
computational fluid dynamics simulations. We extend DVLs to support
large-scale, multi-field adaptive mesh refinement (AMR) data that can be
explored interactively. Our GPU-based system updates the DVL representation
whenever the data or the alpha transfer function changes. We demonstrate and
evaluate our interactive prototype using large AMR volumes from astrophysics
simulations.%
} 

\CCScatlist{
  \CCScatTwelve{Human-centered computing}{Visu\-al\-iza\-tion}{Visu\-al\-iza\-tion application domains}{Visual analytics};
  \CCScatTwelve{Human-centered computing}{Visu\-al\-iza\-tion}{Visu\-al\-iza\-tion application domains}{Scientific visualization}
}

\begin{document}


\firstsection{Introduction} 
\label{sec:intro}

\maketitle

We propose an interactive implementation of
Weissen\-b{\"o}ck~\etal's~\cite{weissenboeck-dvl} \emph{dynamic volume lines}
(DVLs). DVLs visualize ensemble volumes in 1D as a set of polylines. While
ensemble or multi-field volume rendering may suffer from visual clutter and
self-occlusion, DVLs present a viable alternative for visual exploration or can
augment an existing 3D volume visualization. A fundamental problem with DVLs
and similar plots is that many cells of the volume map to only a few pixels of
the output viewport, leading to overdraw. The number of cells can be
several orders of magnitude higher than the number of pixels that the line
segments of the polylines project to, resulting in a linear mapping of cells to
pixels that compresses regions where the data is not as interesting. 

Weissenb{\"o}ck~\etal\ concentrate on the visual analytics aspects of DVLs and
have proven their efficacy to this end, yet the authors' work focused on smaller
structured-regular volumetric data sets (on the order of $64^3$ cells). When
scaling to larger volume sizes and unstructured or hierarchical grid types,
interactively computing DVLs becomes a challenge we address in this paper.

Another aspect that remains unexplored by Weissenb{\"o}ck~\etal's work is that
the spatial arrangement (and hence the local variation) of the volume ensemble
changes when the alpha transfer function of an ensemble member is updated.
However, interactive transfer function updates are an essential aspect of
scientific volume visualization, and the data structures and algorithms
involved in computing and updating DVLs must be carefully chosen not to
prohibit this type of interaction.

We, therefore, concentrate on the performance and interactivity aspects of
computing dynamic volume lines on the GPU. More specifically, we contribute
\begin{itemize}
  \setlength{\itemsep}{0pt}
  \item
an extension of dynamic volume lines for AMR volumes where the cell size
depends on the refinement level,
  \item
a GPU implementation that allows to interactively update the volume lines in
the presence of user-editable transfer functions per ensemble member, and 
  \item
an application that allows interaction with the 3D view and the 1D plot through
brushing and linking.
\end{itemize}
An overview of the visualizations our system supports is given in
\cref{fig:teaser} (here exemplified using a multi-field data set).

\section{Background and Related Work} 
\label{sec:related}
This section reviews related works on large-scale volume visualization and
adaptive mesh refinement (AMR) data. Additionally, we provide a background
summary of the dynamic volume lines method by
Weissenb{\"o}ck~\etal~\cite{weissenboeck-dvl} as our main related work on the
visual analytics side.

\subsection{Large-Scale Volume Data}
Our paper concentrates on large volume data.
While our prototype does include a 3D rendering component, in this section, we
focus on data representation more than on the rendering side.

Although large structured volumes are still commonplace in some
areas~\cite{Hadwiger2012Petascale}, unstructured or hierarchical
representations are ubiquitous in the computational sciences. While
unstructured meshes~\cite{muigg:2011,Sahistan:2021} are pretty standard, many
codes use \emph{adaptive mesh refinement} (AMR)~\cite{Dubey2008} to concentrate
the computation on the relevant regions in space. The resulting data can be
block-structured, overlapping grids, Octrees, or similar hierarchies.

Recent challenges with AMR visualization include smooth interpolation in
3D~\cite{wald:17:AMR}, GPU acceleration structures~\cite{wald-exabricks}, and
time-dependent data~\cite{zellmann2022exajet-animated}. A common approach for
representing AMR data is the one adopted by Wald~\etal~\cite{wald-exabricks},
where AMR cells ``snap'' to the \emph{logical grid}; that hypothetical uniform
grid has a resolution that when resampling the volume, the finest AMR cells
occupy exactly one logical cell. Each AMR cell is unambiguously defined by its
lower corner on the logical grid, and its refinement level $L$. By defining $0$
to denote the finest level, the cell size can be computed as $C_w = 2^L$. This
representation omits the AMR hierarchy itself.

One common way to organize volumes is through space-filling curves (e.g.,
Morton~\cite{zellmann-lbvh,zellmann-binned} or Hilbert
codes~\cite{AmanDG22,Morrical2022QuickClusters}). In 3D rendering, the main
incentive for that, as in the works cited here, is to build acceleration
structures. Generally, space-filling curves cluster cells in 1D that are also
nearby in 3D.

\subsection{Dynamic Volume Lines} 
\label{sec:dvl}

We extend the \emph{dynamic volume lines} (DVL) method by
Weissenb{\"o}ck~\etal~\cite{weissenboeck-dvl} to support large-scale AMR data.
We provide a summary of their method in this section.

DVLs present volumes as 1D plots where the $x$-dimension maps to cell IDs and
the y-dimension maps to intensity. The method is restricted to
structured-regular volumes, i.e., all cells are cubes/voxels of the same size.
One way to assign $x$-values is in a row- or column-major order. This approach
loses spatial locality, as cells are only grouped if they are neighbors on the
same row (or column); yet when they are adjacent in the vertical or depth
direction, they are likely to be further apart in 1D due to the row (or column)
sized stride.

Space-filling curves can provide better proximity-preserving mappings.
Weissenb{\"o}ck~\etal~use Hilbert curves. Cells are represented by their
centroids, which are quantized, e.g., to 20-bit per dimension, so that a 64-bit
bitmask can represent them. The Hilbert codes represent the quantization grid
cell that the centroids map to.

An obvious problem when plotting volumes in 1D is that there are several orders
of magnitude more cells than pixels in the $x$-dimension. Using a linear mapping
is wasteful because it can result in homogeneous or empty regions represented
as straight lines that convey no useful information. 

Weissenb{\"o}ck~\etal's objective is to compare ensembles of volumes.
Interesting features are determined by comparing corresponding points of the
ensemble. The authors propose scaling the cells along the $x$-axis using
per-cell importance. However, their focus is on grayscale volumes and not on
alpha transfer functions as we do. The difference between the maximum and
minimum intensity of the whole ensemble gives the per-cell local variation of
the ensemble:

\begin{equation}
V_{h} = \max_{\forall m \in M}\big(I(m,h)\big) - \min_{\forall m \in M}\big(I(m,h)\big),
\end{equation}
where $h$ is the cell's Hilbert code, $M$ is the ensemble of volumes, and
$I(m,h)$ is the intensity of ensemble member $m$ and the AMR cell that corresponds
to $h$. From that, the authors obtain the local importance scale for the
$x$-coordinate:

\begin{equation}
\label{eq:importance}
f(h) = \Bigg(\frac{V_{h}}{\max\big(V_{h}\big)}\Bigg)^{P}.
\end{equation}
$P$ is a user-defined parameter to control the steepness of the resulting
curve; a minimum importance is enforced (set to $0.025$ by the authors).

Computing a prefix sum of floating point values over the per-cell importance
and quantizing that on the 1D grid given by the width of the plot area provides
$x$ positions to plot the cells as line segments. This nonlinear mapping
compresses regions with low data variation, devoting more screen space to
regions with high data variation.

The concept of exploring spatial or volume data in 1D also inspired other
works. Franke~\etal~\cite{Franke2021}, e.g., use 1D plots to conserve
neighborhood relations of geospatial regions.
Zhou~\etal~\cite{zhou-data-driven-space-filling-curves} use the minimum
spanning tree of a circuit graph over structured-regular volume ensembles to
generate similar 1D plots as the volume lines our paper focuses on. In contrast
to using Hilbert codes directly---and similar to DVLs with their
importance-based nonlinear mapping---Zhou~\etal's approach is also data-driven.

\section{Method} 
\label{sec:method}

We extend Weissenb{\"o}ck~\etal's dynamic volume lines to support multi-field
AMR data. In contrast to Weissenb{\"o}ck~\etal, we assume that the intensities
$I(m,h)$ come from an RGB$\alpha$ transfer function. They do not focus on
interactive parameter updates, such as transfer function, exponent $P$, and
minimum importance; they mention that Hilbert code computation takes several
seconds for $64^3$ cell data sets. While Weissenb{\"o}ck~\etal~focused on
volumes a couple of Megabytes in size, we target Gigabyte-sized data.

\subsection{Extension to AMR} 
\label{sec:importance}

To compute DVLs for AMR data, we need to extend the Hilbert code and
$x$-coordinate generation to support non-uniformly sized cells. First, we note
that Weissenb{\"o}ck~\etal~assume that cells have uniform size; even if the
cells were non-uniform, the Hilbert codes do not reflect this because they only
provide the cells' order and not their spacing. To account for coarser cells to
also span a wider region of space as in a 3D rendering, we need to consider
each cell's size when computing the nonlinear $x$-axis scale. For that, we
extend \cref{eq:importance} to include the AMR cell width as follows:

\begin{equation}
\label{eq:importance2}
f(h) = \Bigg(\frac{V_{h}}{\max\big(V_{h}\big)} 2^{L_{h}}\Bigg)^{P},
\end{equation}
where $L_{h} \in \mathbb{N}_0$ is the AMR level of the cell corresponding to
Hilbert code $h$, and $0$ is the finest level. Another sensible choice would be
to not scale the cells by their (uniform) width but by their volume.

We deliberately apply the cell-size dependent scale when computing the
importance, not the $x$-positions themselves, because the order of operations
is not commutative. We compute the floating point prefix sum over the
importance values:

\begin{equation}
\label{eq:prefix-sum}
F(h) = \sum_{i=0}^{h}f(i),
\end{equation}
to obtain $x$-coordinates $xf_1 = \frac{F(h-1)}{F(\max{(h)})} \times W$, $xf_2
= \frac{F(h)}{F(\max{(h)})} \times W$ for plots of size $W$ pixels.

\begin{figure*}[tb]
\centering
\includegraphics[width=0.95\linewidth]{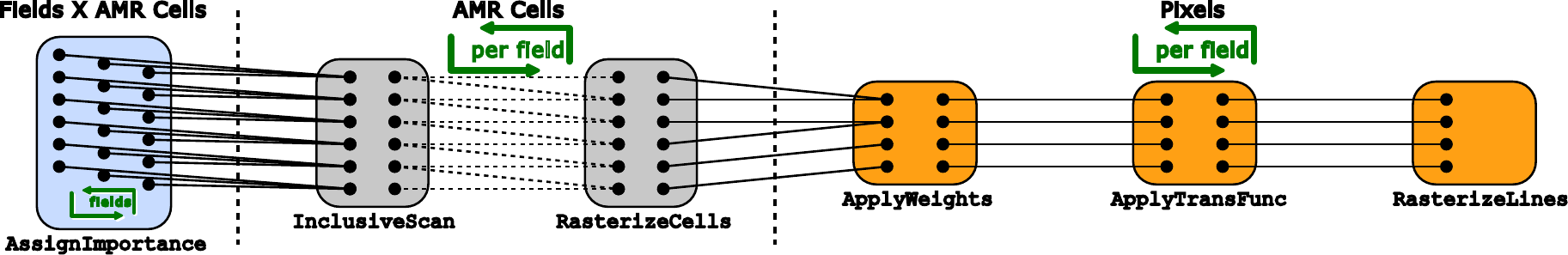}
\vspace{-1.2em}
\caption{\label{fig:kernels}%
CUDA kernels executed on volume line updates. Blue indicates kernels processing
each AMR cell and scalar field. Gray indicates kernel execution per AMR cell
(these get executed \emph{per} field). Orange boxes indicate kernels executed
per (horizontal) raster point (also per field). As a rule of thumb, the number
of work items generally decreases from left to right. Transitioning from
``Fields $\times$ AMR Cells'' to ``AMR Cells'' is realized using loops but
requires no synchronization. Transitioning from ``AMR Cells'' to ``Pixels''
uses atomic operations on CUDA global memory.
}
\vspace{-2.0em}
\end{figure*}

\subsection{Projecting Cells to 1D} 
\label{sec:rasterize}
We are now able to compute pairs $(xf_1,xf_2)$ of $x$-coordinates per AMR cell;
we note that these $x$-coordinates have sub-pixel accuracy, and although we
apply the nonlinear mapping using cumulative importance, in general, a
multitude of $x$-coordinates will project to single pixels. We now discuss how
to map these coordinate pairs to obtain $x$-coordinates per horizontal pixel $x
\in W$ and how to obtain ``$y$-values'' for these. Given such pairs $(x,y)$, we
can draw line strips with control points per pixel in the $x$ dimension.

For that, we create a set of $W$ bins---one for each pixel in the plot's
$x$-dimension---whose values we initialize to $0$. We then project the pairs
$(xf_1,xf_2)$ to integer coordinates in the range $[0,W-1]$, iterate over
these, and increase the overlapping bins by the value of the corresponding AMR
cell. We also maintain a per-bin counter that we increment whenever we increase
the bin value. After all the pairs $(xf_1,xf_2)$ are processed, we iterate over
the bins and divide each by its bin counter, obtaining the average intensity
value of all the AMR cells that project to the bin. 

This procedure borrows from the basis function method by
Wald~\etal~\cite{wald:17:AMR}, only that we use a box-shaped basis function
instead of the tent-shaped basis used by Wald. Pairs $(xf_1,xf_2)$ that span
multiple bins (hence multiple pixels in the $x$-dimension) will noticeably turn
the plot into a step function. As for our data, many line segments will map to
single bins only; we do not consider this to be an issue, and we here choose
simplicity over generality.

\subsection{Interactive Transfer Function Updates} 
\label{sec:tf}
We extend Weissenb{\"o}ck~\etal's method to support interactive RGB$\alpha$
transfer functions that apply to 1D and 3D rendering alike. The requirement
that transfer function updates be interactive implies that (re)generating
volume lines must also be interactive.

Transfer functions are applied on two occasions: once when computing the
importance (cf.\ \cref{sec:importance}), which requires intensities from the
transfer function, and once per bin, after dividing the bin values by their
basis weights. We normalize the input (field) intensity and compute RGB$\alpha$
values to apply the transfer function. The alpha value determines the height of
the bins and y-values of the polyline at these positions. The RGB value (or,
alternatively, a uniform color from a global map) is used to colorize the
polylines.

\subsection{Computing the Maximum Local Ensemble Variation} 
\label{sec:variation}
The term $\max\big(V_{h}\big)$ from \cref{eq:importance}, the maximum of the
local ensemble variations for each Hilbert code $h$, needs to be recomputed
whenever an ensemble member's transfer function changes. This can be
implemented on GPUs using parallel \texttt{reduce} over all cells or a kernel
atomically updating a single value in GPU main memory.

To avoid this costly operation, we compute this value using the transfer
functions and field data ranges, resulting in a more conservative, yet in
practice very close approximation to $\max\big(V_{h}\big)$.


We compute ranges $[i_m,j_m]$ where $i_m,j_m \in [0,N-1]$ and $i_m \le j_m$ for
each ensemble member $m \in M$ and transfer function size $N$. These allow us
to iterate only over the transfer function values present in the data. That
way, we compute the global range $[i,j]$: 
\begin{equation}
i = \min_{\forall m \in M}(i_m), j = \max_{\forall m \in M}(j_m),
\end{equation}
\noindent and from that
\begin{equation}
\label{eq:va}
V_{a} = \max_{\forall m \in M}\big(\mathrm{A}(m,a)\big) - \min_{\forall m \in M}\big(\mathrm{A}(m,a)\big)~\forall a \in [i,j]
\end{equation}
\noindent as an approximation to $V_h$. In \cref{eq:va}, the term $\mathrm{A}$ denotes a lookup to the transfer function to retrieve the alpha value.

\subsection{Brushing and Linking} 
\label{sec:brushing}
We connect the DVL and 3D views using \emph{brushing and linking}. When
selecting regions of interest (ROIs) in the 1D plot, the corresponding cells in
the 3D view are highlighted (cf.\ \cref{fig:gui}). We use the ROIs' first and
last Hilbert codes as selection ranges for that. In the 3D shader, the ROIs
also manifest as Hilbert codes and not as lists of cells in world space. This
compact representation comes at the expense of transforming the center of the
cell that we are sampling to Hilbert space to test it against the ROIs.

In the case of structured-regular volumes, finding the cell bounds is simple;
since the cells have the same size, the corresponding Hilbert code implicitly
allows us to derive their (uniform) size. For AMR data, the cell centroid's
Hilbert code is not sufficient; instead, when we check if a sample falls inside
an ROI, we must explicitly locate cells to know the ROI's exact size.

\section{GPU Implementation with CUDA} 
\label{sec:cuda}
We implement what we call \emph{interactive volume lines} (IVLs)---the
GPU-accelerated version of dynamic volume lines---using NVIDIA CUDA. The CUDA
kernels involved are shown in \cref{fig:kernels}, corresponding to the
algorithm phases from \cref{sec:method}.
We now discuss how to assemble these building blocks and identify their
bottlenecks. The Hilbert codes and the order of the 1D AMR cells never change
once they are established. What does change in user interaction is the spacing
between cells that we need to recompute interactively.

We compute the term $V_{a}$ from \cref{eq:va} on the CPU since the transfer
function color map and other parameters are passed to our application through
the host in any case and because the amount of computation needed does not
necessitate running this operation in parallel. With $V_{a}$ computed,
execution transitions to the GPU.

A kernel with one thread per AMR cell (\texttt{AssignImportance} in
\cref{fig:kernels}) computes the importance from \cref{sec:importance}. The
kernel loops over each field, assigning each cell its transfer function and
cell-size-dependent importance from \cref{eq:importance2}. This requires
exclusive read and write accesses (EREW) only. For \cref{eq:prefix-sum}, we use
the \texttt{InclusiveSum} algorithm from the CUB library
(\texttt{ExclusiveScan} in \cref{fig:kernels}). GPU prefix sums can be realized
with EREW accesses~\cite{harris}.

We run another kernel with one thread per cell (\texttt{RasterizeCells} in
\cref{fig:kernels}) to determine the subpixel $x$-coordinates $xf_1$,$xf_2$
from the prefix sum array and update the bins and weights (cf.\
\cref{sec:rasterize}). The number of bins is much smaller than the number of
AMR cells for typical data sizes. We use CUDA atomic operations for the
projection; hence, this kernel is not EREW. Due to the input size and the
atomics, this is the most costly kernel of the algorithm.

The \texttt{ApplyWeights} and \texttt{ApplyTransFunc} kernels (cf.\
\cref{fig:kernels}) are EREW. They divide the bin values by their weight (cf.\
\cref{sec:rasterize}) and apply the transfer function to obtain
$y$-coordinates. Both kernels use $W$ (number of bins) threads. We finally run
a kernel (\texttt{RasterizeLines} in \cref{fig:kernels}) rasterizing the
polylines using CUDA surfaces. We implemented modes to draw the IVLs as
polylines (\cref{fig:teaser}b) or as bar charts (\cref{fig:teaser}c).

\section{Prototypical User Interface}

\begin{figure}[tb]
\centering
\begin{tikzpicture}
 \node[anchor=south west,inner sep=0] (image) at (0,0) {\includegraphics[width=0.999\columnwidth]{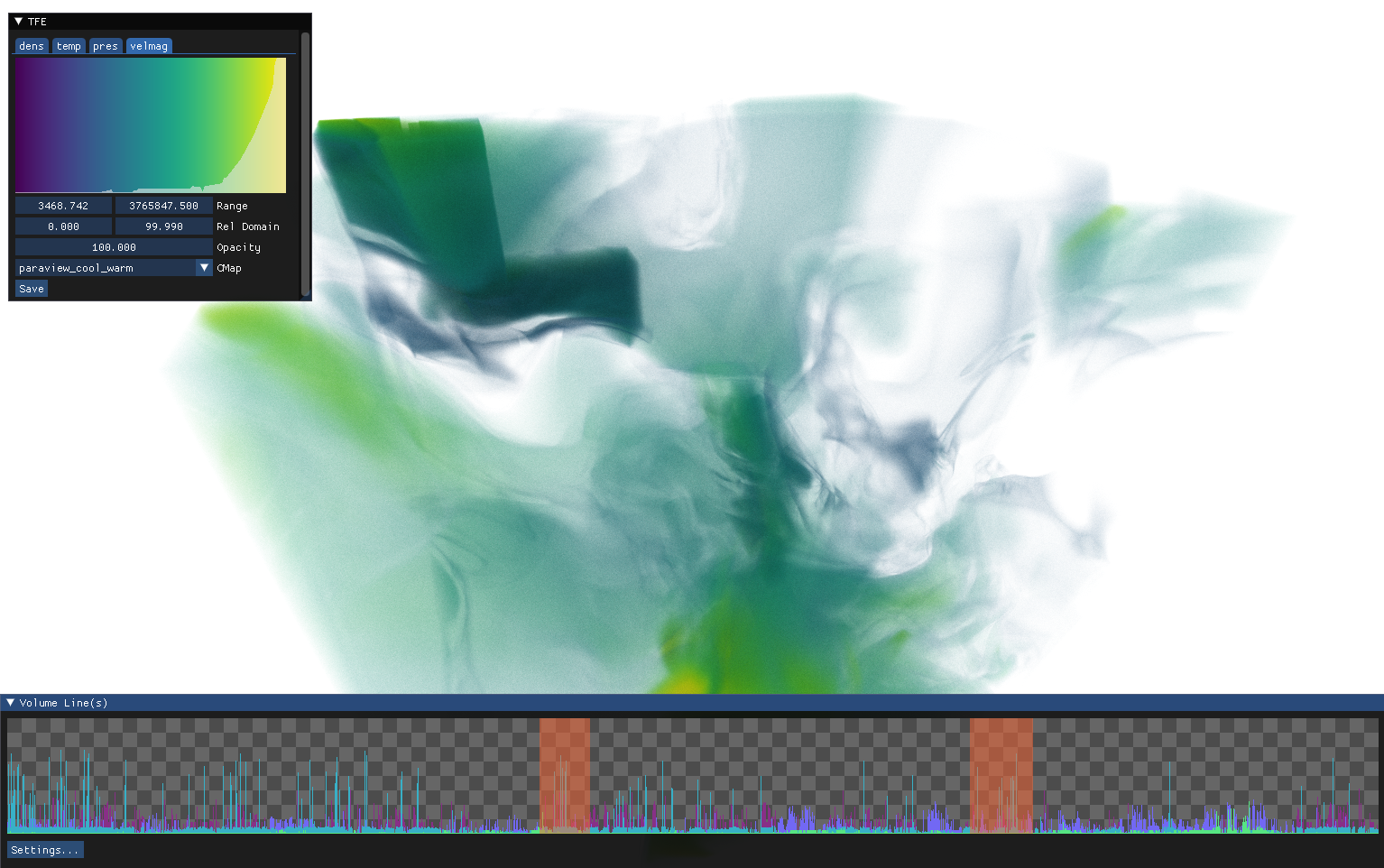}};
 \node[shape=circle,draw,fill=white,minimum size=14pt,inner sep=0, line width=1pt] (char) at (0.3,5) {a};
 \node[shape=circle,draw,fill=white,minimum size=14pt,inner sep=0, line width=1pt] (char) at (0.3,1) {b};
 \node[shape=circle,draw,fill=white,minimum size=14pt,inner sep=0, line width=1pt] (char) at (2.9,0.6) {c-i};
 \node[shape=circle,draw,fill=white,minimum size=14pt,inner sep=0, line width=1pt] (char) at (5.5,0.6) {d-i};
 \node[shape=circle,draw,fill=white,minimum size=14pt,inner sep=0, line width=1pt] (char) at (3.4,4.4) {c-ii};
 \node[shape=circle,draw,fill=white,minimum size=14pt,inner sep=0, line width=1pt] (char) at (5.5,2.6) {d-ii};
\end{tikzpicture}
\vspace{-2.2em}
\caption{\label{fig:gui}%
Our prototype, used with brushing and linking. Interactive transfer function
updates~(a) affect 3D rendering and the IVL plots shown in~(b). Here we
selected the two regions (c-i) and (d-i) in the IVL plot that are connected to
the 3D view through brushing and linking and are highlighted using dimming
((c-ii) and (d-ii)).
}
\vspace{-1em}
\end{figure}
To implement a prototype, we started with the open-source code of
Zellmann~\etal~\cite{stitcher} and added multi-field support and a 1D user
interface with Dear~ImGUI~\cite{imgui}. We show the user interface
demonstrating brushing and linking in \cref{fig:gui}. We note that the
application is at an early stage and does not implement all the features
proposed by Weissenb{\"o}ck~\etal~\cite{weissenboeck-dvl}, such as mouse-over
handling or support for the histogram heatmap and functional boxplots.
Fundamentally, the computations required are of the same order as those
required for the IVLs, so implementing them remains an engineering exercise.

\section{Evaluation} 

\begin{table}[t]
  \centering
  \caption{\label{tab:kernels}%
  Execution times for the CUDA kernels from \cref{fig:kernels} and the data set from \cref{fig:teaser}.
  \vspace{-0.6em}
  }
  \scalebox{0.9}{
  \begin{tabular}{l|c}
    \toprule
     Kernel    &  Execution Time (ms) \\
    \midrule
    \texttt{AssignImportance} & 3 \\
    \texttt{InclusiveScan}    & 0.4 \\
    \texttt{RasterizeCells}   & 55 \\
    \texttt{ApplyWeights}     & $<~$0.01\\
    \texttt{ApplyTransFunc}   & $<~$0.01\\
    \texttt{RasterizeLines}   & 0.2 \\
    \bottomrule
  \end{tabular}
  }
  \vspace{-2.2em}
\end{table}

\label{sec:eval}
We evaluate our method on an Intel Xeon system with 64~GB RAM and an NVIDIA
A6000 GPU. We use the \texttt{dens}, \texttt{temp}, \texttt{pres}, and
\texttt{velmag} fields of the Molecular Cloud data set by
Seifried~\etal~\cite{seifried_2017}, which was simulated with
FLASH~\cite{Dubey2008}. Technically, the data is multi-field and not an
ensemble; the fields are correlated. The data set spans four AMR levels with a
total of 35.8~M cells. If not noted otherwise, we set $P=1$ (cf.\
\cref{eq:importance2}) and the minimum importance to $0.025$. GPU memory for
the whole data set---including auxiliary data (our application uses OptiX to
accelerate 3D rendering)---is reported by \texttt{nvidia-smi} to amount to
3.6~GB.

We present kernel execution times in ms.\ in \cref{tab:kernels}. The algorithm
is bottlenecked by the \texttt{RasterizeCells} kernel (projection from cells to
bins using atomics), executed once per field.

\begin{figure}[t]
\centering
  \includegraphics[width=0.49\columnwidth]{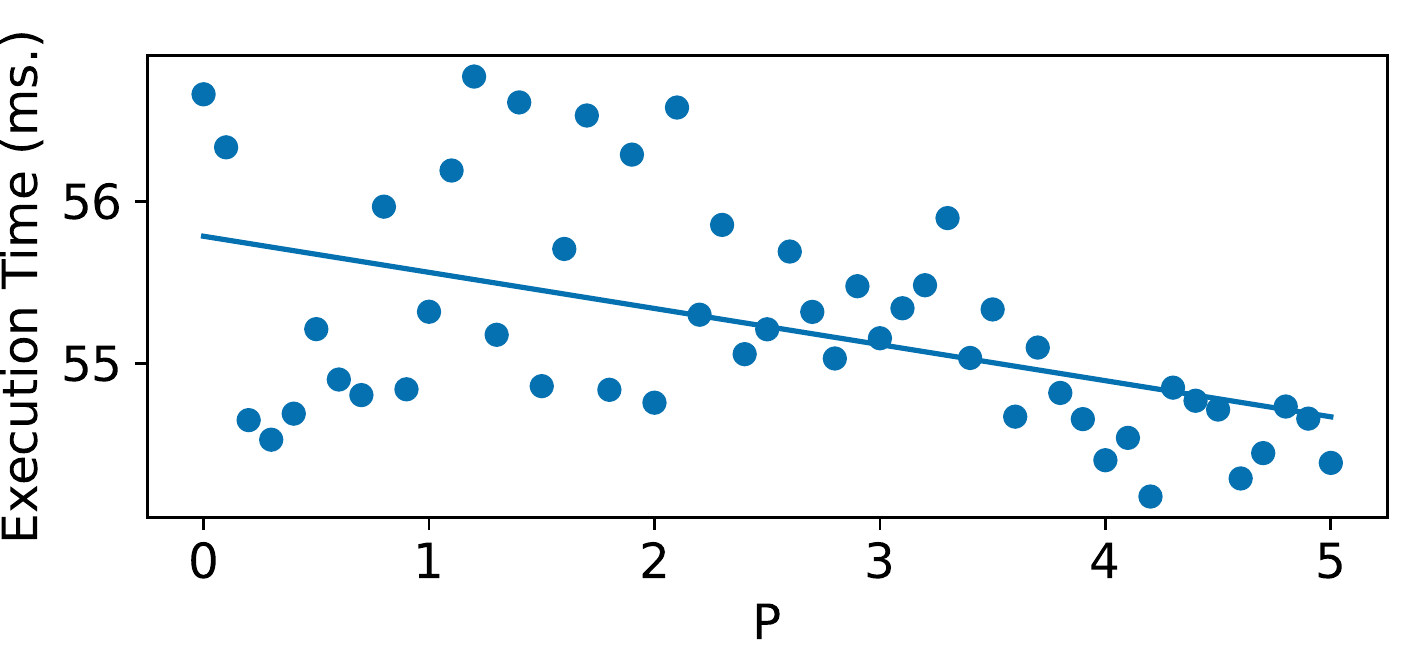}
  \includegraphics[width=0.49\columnwidth]{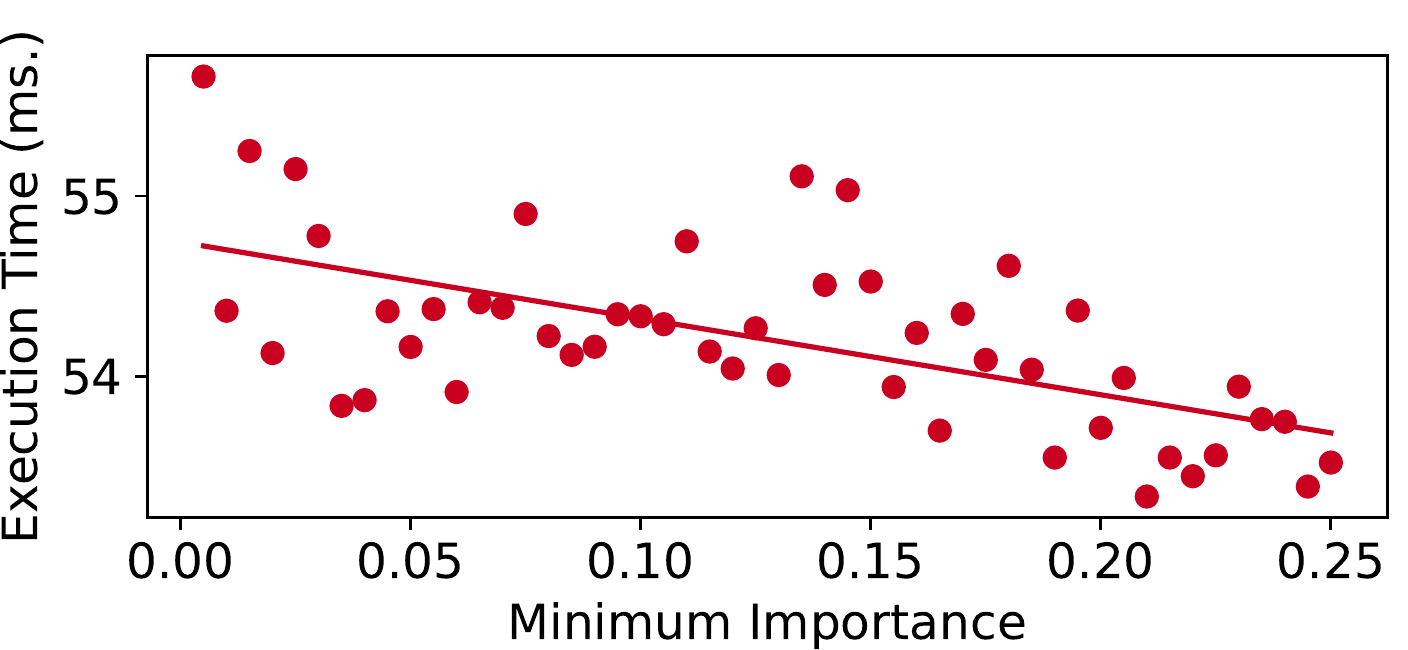}
\vspace{-1.5em}
\caption{\label{fig:timings}%
Execution times of the \texttt{RasterizeCells} kernel (cf.\ \cref{fig:kernels})
as a function of parameter $P$ (left) and of minimum importance (right).
}
\vspace{-1.5em}
\end{figure}

We test if the performance of the \texttt{RasterizeCells} kernel
depends on the input parameters. While keeping the other parameters fixed, we
vary $P$ in $0.0-5.0$ (\cref{fig:timings}, left), and the minimum importance in
$0.0-0.25$ (\cref{fig:timings}, right). We observe that there \emph{is} a
measurable, yet very subtle (1-2\%) trend that parameters that favor higher IVL
compression lead to slightly faster execution times.

\begin{figure}[t]
\centering
  \includegraphics[width=0.49\columnwidth]{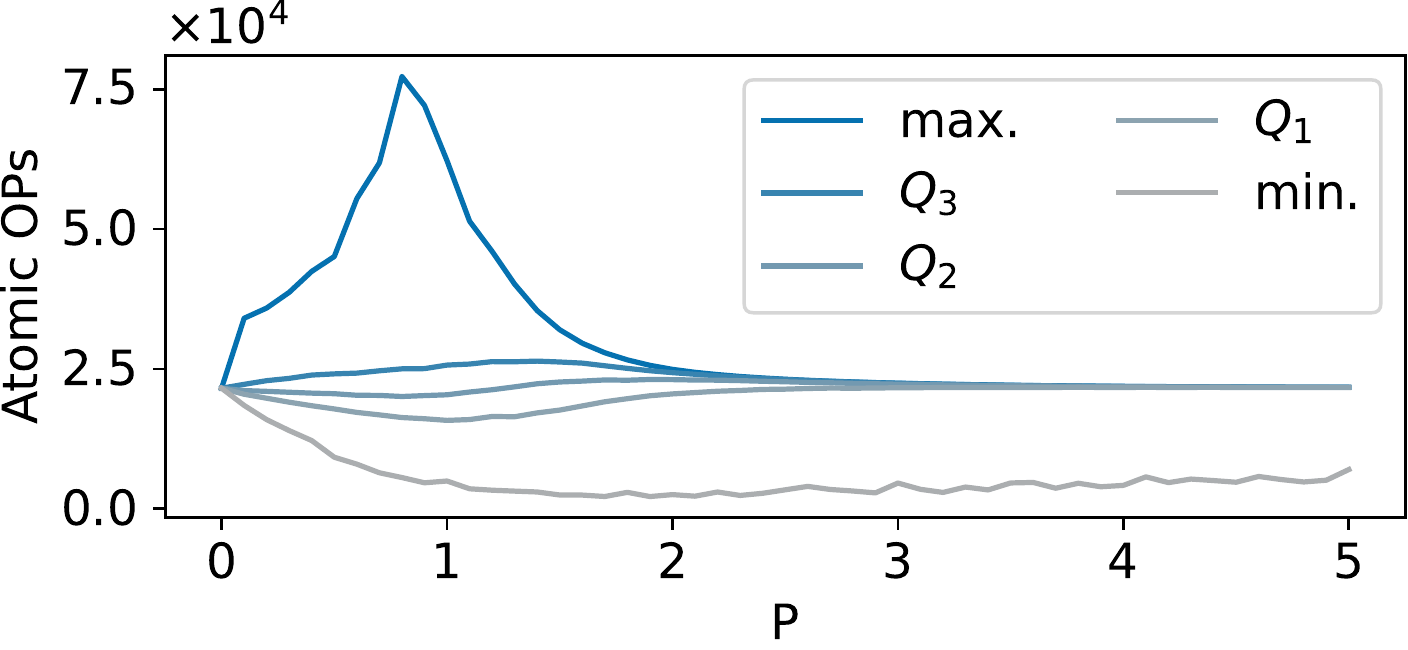}
  \includegraphics[width=0.49\columnwidth]{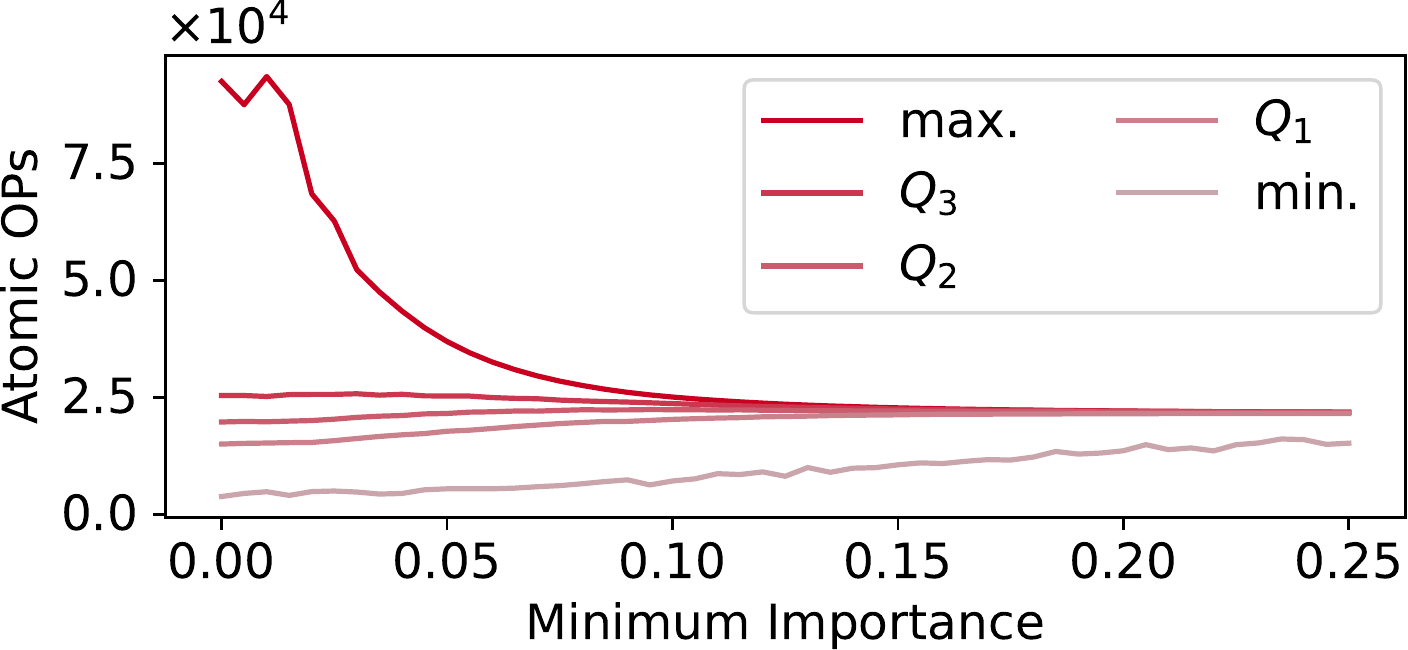}
\vspace{-1.5em}
\caption{\label{fig:atomics}%
Atomic operations performed per bin by the \texttt{RasterizeCells} kernel (cf.\
\cref{fig:kernels}) as a function of parameter $P$ (left) and of minimum
importance (right). We show the minimum and maximum count as well as the
quartiles across bins.
}
\end{figure}
\begin{figure}[!tb]
\centering
\vspace{-1em}
\includegraphics[width=0.49\columnwidth]{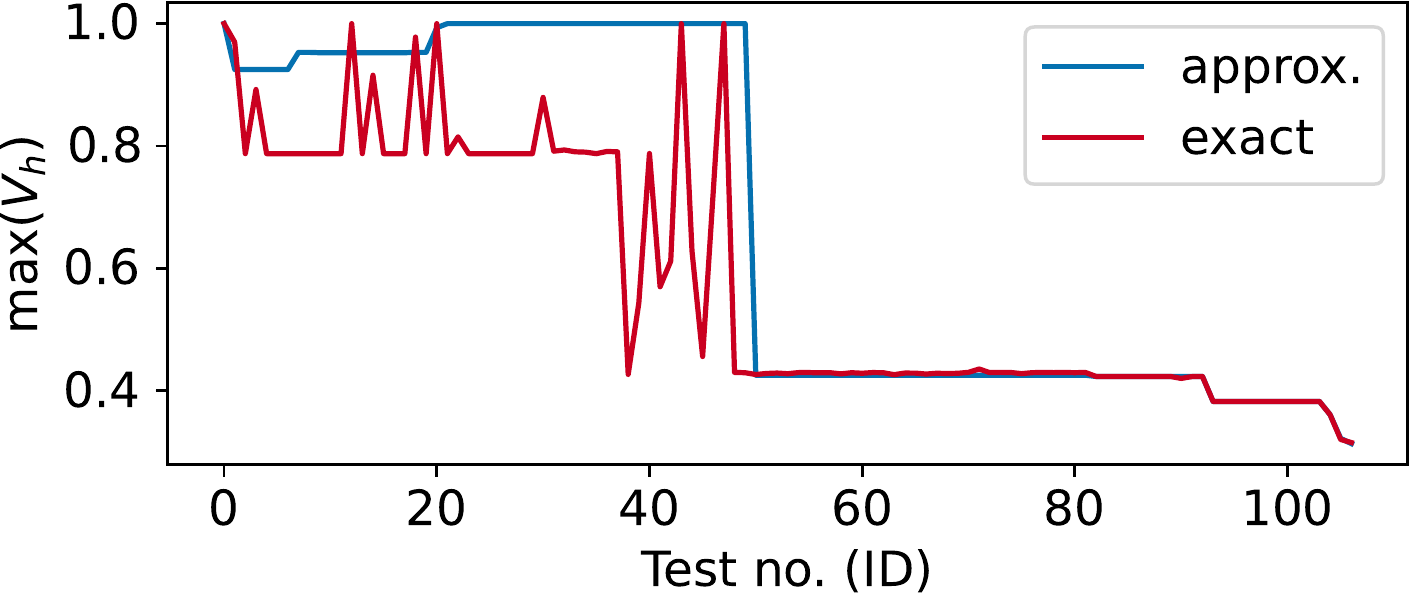}
\vspace{-1.5em}
\caption{\label{fig:vh-difference}%
Comparison of exact $\max(V_h)$ computation (red) vs.\ our approximation from
\cref{sec:variation} (blue).
}
\vspace{-2em}
\end{figure}

We count the \texttt{atomicAdd}'s to determine their costs, and report the bin
minima and maxima as well as quartiles for the same test from before (cf.\
\cref{fig:atomics}). We observe that when $P$ and the minimum importance
increase, the number of \texttt{atomicAdd}'s per bin becomes more uniform. We
never encountered single threads that run extraordinarily long. In fact, during
development, we tested if parallelizing the kernel using one
\texttt{cudaStream} per field lowered total execution time, but found that
it fully occupies the GPU at all times.

We finally compute the difference between the exact $\max(V_h)$ used by
Weissenb{\"o}ck~\etal's~\cite{weissenboeck-dvl} vs.\ our approximation from
\cref{sec:variation}, using a sweep of $107$ randomly chosen different transfer
function configurations. We report the results in \cref{fig:vh-difference}.


\section{Conclusions and Future Work}
\label{sec:summary}
We presented \emph{interactive} volume lines, a high-performance variant of
Weissenb{\"o}ck~\etal's~\cite{weissenboeck-dvl} dynamic volume lines. For
non-trivial volume data (e.g., unstructured or AMR), visual analytics performed on
the whole set of cells becomes a complex data handling task. We presented a
carefully crafted GPU implementation of this algorithm.

For large-scale volumes, as demonstrated in this paper, it does matter if an
operation is performed on the whole set of cells, on bins of a 1D grid in
screenspace, or on the RGB$\alpha$ transfer function array, and which type of
write accesses are performed. The GPU control flow we eventually developed
resulted in a balance between faithfully recreating the algorithm and avoiding
severe implementation bottlenecks.

Our paper points to future work. One conceivable optimization is to make the
kernel that the algorithm is bottlenecked on become hierarchical over the input
cells to perform the projection in more local memory regions. This is one of
many interesting similarities between optimizations used for 3D volume
rendering (e.g., acceleration structures like the one by
Wald~\etal~\cite{wald-exabricks}) and optimizations that apply to IVLs. In
fact, we found that on the engineering side, similar abstractions apply in 1D
and 3D alike (e.g., basis functions used for interpolation, 1D bins that
resemble pixels, etc.).

One less obvious future work lies in the algorithm's GPU memory consumption: 1D
and 3D rendering share neither the data itself nor other auxiliary data
structures that accelerate rendering. Instead, the whole input data is
replicated in memory. It would be interesting to explore if the two rendering
modes could share some or even all the data to overcome this limitation.

Another open question (that this paper does not seek to answer) is how useful
volume lines are for large volumes; our intuition is that they work best for
smaller data, to reduce noise and outliers in the 1D plots. A formal evaluation
is out of scope here. We note though that an obvious extension would be a zoom
interaction that, when selecting a ROI through brushing and linking, also zooms
in on the 1D plot and devotes more screen space to the selected AMR cells.





\balance

\acknowledgments{This work was supported by the Deutsche Forschungsgemeinschaft (DFG, German Research Foundation)---grant no.~456842964. The TAC Molecular Cloud is courtesy of Daniel Seifried. We are grateful to NVIDIA, who kindly provided us with the hardware we used for the evaluation.}

\bibliographystyle{abbrv-doi}

\bibliography{template}

\balance

\end{document}